\documentclass[amssymb,amsmath,twocolumn,prx]{revtex4}

\usepackage{color,graphicx,bm}
\usepackage{overpic}
\usepackage{upgreek}

\newcommand{\ua}{\uparrow}
\newcommand{\da}{\downarrow}

\def\k{{\bf k}}
\def\q{{\bf q}}
\def\b #1{{\bf #1}}

\begin{document}

\title{Route to Topological Superconductivity via Magnetic Field Rotation\\[0mm]
}

\author{
Florian Loder$^{\,1,2}$, Arno P. Kampf$^{\,2}$, and Thilo Kopp$^{\,1}$
\vspace{0,3cm}}

\affiliation{Center for Electronic Correlations and Magnetism, $^1$Experimental Physics VI, $^2$Theoretical Physics III,\\ 
Institute of Physics, University of Augsburg, 86135 Augsburg, Germany}

\date{\today}

\begin{abstract}
The verification of topological superconductivity has become a major experimental challenge. Apart from the very few spin-triplet superconductors with $p$-wave pairing symmetry, another candidate system is a conventional, two-dimensional (2D) $s$-wave superconductor in a magnetic field with a sufficiently strong Rashba spin-orbit coupling. Typically, the required magnetic field to convert the superconductor into a topologically non-trivial state is however by far larger than the upper critical field $H_{\text{c2}}$, which excludes its realization. In this article, we argue that this problem can be overcome by rotating the magnetic field into the superconducting plane. We explore the character of the superconducting state upon changing the strength and the orientation of the magnetic field and show that a topological state, established for a sufficiently strong out-of-plane magnetic field, indeed extends to an in-plane field orientation. We present a three-band model applicable to the superconducting interface between LaAlO$_3$ and SrTiO$_3$, which should fulfil the necessary conditions to realize a topological superconductor.
\end{abstract}


\maketitle

\section{Introduction}
While topologically non-trivial superconducting (SC) states have been established  theoretically in numerous systems~\cite{sato1:09,sato2:09,sau:10,alicea:10,qi:11,bernevig:13}, an experimental verification of such a state is still awaited. This is largely a consequence of the required conditions, which tend to counteract superconductivity itself. A topologically non-trivial state is generally described by a non-zero momentum space Berry phase $\gamma=2\pi C$ with an integer $C$ whenever there is an energy gap separating occupied from unoccupied states~\cite{qi:11}. The superconducting state can acquire a finite Berry phase through a chiral order parameter, and also via gapping a chiral normal-metal state upon entering a conventional SC state. Examples of the former case are selected spin-triplet states, e.g., the B-phase of superfluid $^3$He~\cite{qi:11} and most likely the superconducting phase of Sr$_2$RuO$_4$~\cite{sigrist:99,mackenzie:03}. Very recently, a similar topological character was also proposed for the superconducting state in strongly underdoped cuprates~\cite{lu:14} in which a gap exists even along the nodal direction of a $d$-wave order parameter~\cite{vishik:12}. These proposals are built on states of matter, where the topological nature is an intrinsic property. The chiral order parameter however requires a very special pairing interaction; the $p$-wave states are rare in nature and pose considerable experimental challenges.

On the other hand, if the material provides a chiral band structure by itself, a conventional BCS superconductor with an $s$-wave order parameter can be topologically non-trivial as well. Most often discussed is an $s$-wave superconductor with a Rashba type spin-orbit coupling (SOC) in two dimensions~\cite{sato2:09}. Rashba SOC generates two bands with opposite chirality~\cite{loder:13}. Therefore, in order to reach a state with a finite overall Berry phase, an additional Zeeman field is needed which is strong enough to depopulate one of the SOC split bands. The topological character of the resulting SC state is equivalent to the quantum-Hall state. Such states are classified by a topological invariant, the so-called TKNN integer $C$ (after Thouless, Kohmoto, Nightingale, and Nijs)~\cite{thouless:82}. If the magnetic field is perpendicular to the plane of the 2D superconductor, the minimal Zeeman splitting required to reach the topological phase is $\mu_\text{B}H_\text{t}=\sqrt{\epsilon_0^2+\Delta^2}$, where $\epsilon_0$ measures the distance of the band energy at $\k=\bm0$ to the Fermi energy, and $\Delta$ is the SC energy gap~\cite{sato2:09}. The obstacle for realizing this topological state experimentally is to find a system which remains superconducting in the required high magnetic fields.
Suggested model systems are, e.g., neutral ultra-cold atoms in an optical trap~\cite{sato2:09}, or heterostructures where Cooper pairs are induced through the proximity effect~\cite{kitaev:01,sau:10,lutchyn:10,oreg:10,zyuzin:13}.

The problem of realizing the topological $s$-wave state has two distinct aspects: ($i$) $\mu_\text{B}H_\text{t}$ must be larger than $\Delta$. While the presence of the Rashba SOC allows in principle $s$-wave superconductivity in a Zeeman field larger than $\Delta$, the orbital critical field $H_{\text c2}$ is typically much smaller. ($ii$) The superconductor must have $\epsilon_0$ smaller than the Zeeman splitting. This requires a low band filling and, therefore, superconductivity must be stabilized by yet another band with larger filling. In this article, we address both of these aspects and demonstrate that the problems can be overcome in real solid-state systems.

\begin{figure*}[t!]
\centering
\vspace{3mm}
\begin{overpic}
[width=1.75\columnwidth]{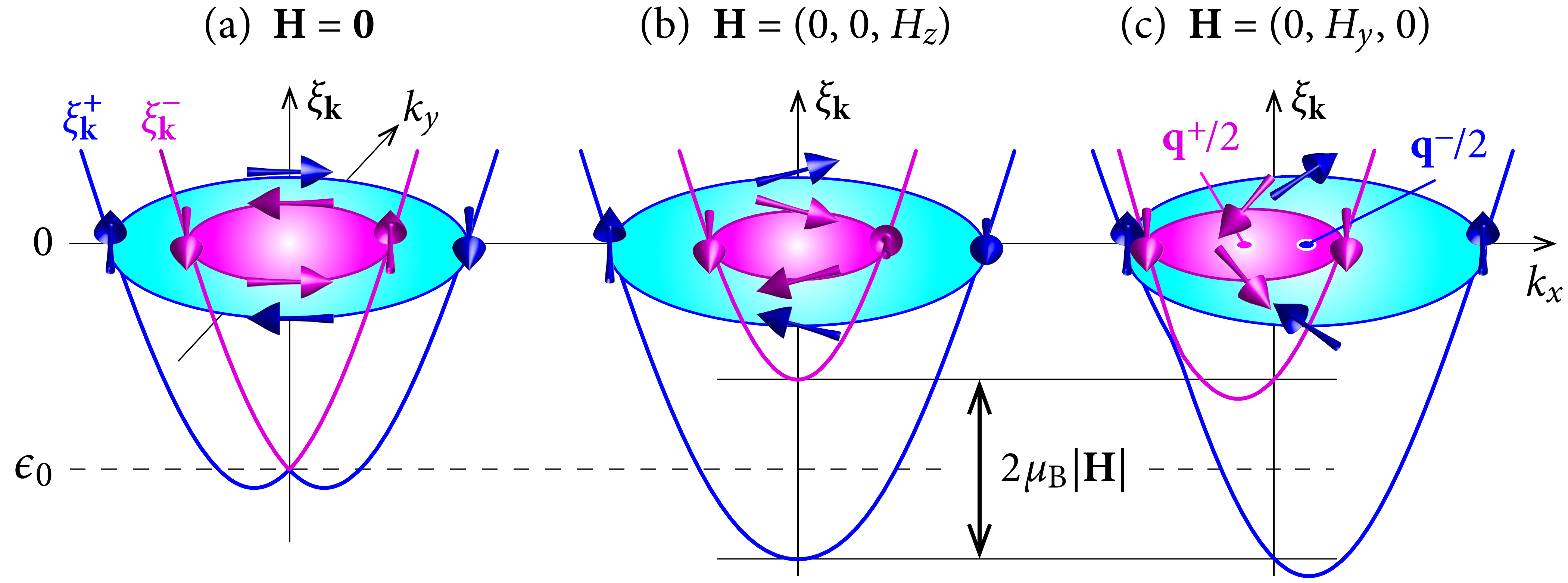}
%\put(2,31){(a)}
%\put(34,31){(b)}
%\put(67.5,31){(c)}
\end{overpic}
\vspace{-2mm}
\caption{Energy bands $\xi_\k^+$ (pink) and $\xi_\k^-$ (blue) and Fermi surfaces with $\alpha>0$ and magnetic filed $|\b H|<H_\text{t}$. (a) For $\b H=\bm0$, the two bands touch at $\k=\bm0$. (b, c) For $|\b H|>0$, the band splitting at $\k=\bm0$ is equal to the Zeeman splitting $2\mu_\text{B}|\b H|$. The centers of the shifted Fermi surfaces in (c) are at the momenta momenta $\q^+/2=(q^+/2,0)$ and $\q^-/2=(q^-/2,0)$, respectively. Although $q^+\approx -q^-$, their absolute values are in general different. Note that the Fermi energy is somewhat larger than $|\epsilon_0|$ because of the SOC induced band splitting.
}
\label{Fig1}
\end{figure*}

A simple way to circumvent the orbital critical field $H_\text{c2}$ is to rotate the magnetic field into the plane of the 2D superconductor. The in-plane field however leads to an unusual type of pairing. In the presence of Rashba SOC, an in-plane magnetic field shifts the Fermi surfaces out of the Brillouin-zone center (cf. Fig.~\ref{Fig1}), and the electron pairs thereby acquire a finite center-of-mass momentum (COMM)~\cite{loder:13}. Edge states in an in-plane magnetic field have recently been investigated for $p$-wave superconductors, but with zero COMM~\cite{sato1:09,wong:13}. As we show here, the inclusion of a finite COMM in such a field geometry is indispensable for the discussion of topology. Specifically, we analyze the topological properties of an $s$-wave superconductor under rotation of the magnetic field within a fully self-consistent treatment of the SC order parameter. It is verified that the topological state reached in out-of-plane fields indeed persists to in-plane field orientations, if the COMM is appropriately chosen to minimize the free energy. For in-plane fields the energy gap closes, accompanied by a topological transition. Nevertheless, chiral edge modes remain even for a regime with a closed gap.

We discuss the experimental realizability of a topological $s$-wave superconductor in a nearly in-plane magnetic field. As a candidate system, which can possibly fulfil the required conditions, we consider the metallic LaAlO$_3$-SrTiO$_3$ (LAO-STO) interface~\cite{ohtomo:04,thiel:06}. 
For this system, several models are proposed for a topologically non-trivial superconducting state, which rely on an unconventional order parameter~\cite{fidkowski:11,fidkowski:13,scheurer:14}. 
Assuming instead an $s$-wave pairing state, we demonstrate that a multi-band model involving the titanium $t_{2g}$ orbitals allows for a topologically non-trivial superconductor in a realistic parameter regime for the LAO-STO interface. We suggest that it may be achieved with the currently used experimental setups.

\section{Topological Characterization}

In order to investigate the magnetic-field dependence of an $s$-wave superconductor with Rashba SOC in transparently simple terms, we use a one-band tight-binding model on a square lattice in the $x$-$y$-plane at zero temperature.
In our analysis of the topology upon rotating the magnetic field $\b H$ into the plane, we include the Zeeman coupling of the electrons to the magnetic field, but neglect the orbital coupling. This approximation is well justified for the nearly in-plane field orientation on which we focus here; but orbital effects are necessarily important for the superconducting state in an out-of-plane magnetic field.

The Rashba SOC and the Zeeman coupling to the magnetic field $\b H$ are combined into
\begin{align}
{\cal H}_{\rm S}=\sum_{\k,s}\b h_\k\cdot\bm\sigma_{ss'}\,c^\dag_{\k,s}c_{\k,-s}
\label{g1}
\end{align}
with $s=\pm1$, the Bloch vector $\b h_\k=\alpha\b g_\k+\mu_\text{B}\b H$, and $\b g_\k=(\sin k_y, -\sin k_x,0)$; $\bm\sigma$ is the vector with the Pauli matrices as components. The strength of the Rashba SOC $\alpha$ derives originally from the Dirac Hamiltonian, but may have other sources in multi-band systems (see Sec.~\ref{sec4}). Diagonalizing the kinetic energy together with ${\cal H}_\text{S}$ gives the two chiral energy bands
$\xi^\pm_{\k}=\epsilon_\k\pm|\b h_\k|$, where $\epsilon_\k=-2t(\cos k_x+\cos k_y)-\mu$ with the nearest-neighbor hopping amplitude $t$ and the chemical potential $\mu$ (thus $\epsilon_0=-4t-\mu$).
In these bands, the spin is either parallel or antiparallel to $\b h_\k$ and has a component which rotates either counter-clockwise or clockwise upon circulating the Fermi surfaces~(see Fig.~\ref{Fig1}).

\subsection{Out-of-plane magnetic field}

For an out-of-plane magnetic field with $H_x=H_y=0$, the topological properties of the superconducting state are readily established (see e.g.~Ref.~[\onlinecite{sato2:09}]).
Its Hamiltonian ${\cal H}=\sum_\k\b C^\dag_\k H_\k\b C_\k$ is represented by the 4$\times$4 matrix
\begin{align}
H_\k=\begin{pmatrix}\epsilon_\k\sigma^0+\b h_\k\cdot\bm\sigma&i\sigma^y\Delta\cr-i\sigma^y\Delta^{\!*}&-\epsilon_{-\k}\sigma^0-\b h^*_{-\k}\cdot\bm\sigma^*\end{pmatrix}
\label{g5}
\end{align}
with $\b C^\top_\k=(c_{\k,\ua},c_{\k,\da},c^\dag_{-\k,\ua},c^\dag_{-\k,\da})$  and $\sigma^0$ is the 2$\times$2 unit matrix; the pairing field $\Delta$ is calculated self-consistently from Eq.~(\ref{g4}) (with $\q=\bm0$).
The four eigenenergies $E_{\k,n}$ obtained from diagonalizing (\ref{g5}) are generally the solutions of a 4th order polynomial, but simplify to
\begin{align}
E_{\bm0,n}=\pm\sqrt{\epsilon_{0}^2+\Delta^2}\pm\mu_\text{B}|\b H|,
\label{g6}
\end{align}
for $k_x=k_y=0$, since $\b g_{\k=\bm0}=\bm0$. The number $n$ labels the four combinations of the plus and minus signs. It follows that the energy gap closes at $\k=\bm0$ for $\mu_\text{B}^2H^2_z=\epsilon_{0}^2+\Delta^2\equiv\mu_\text{B}^2 H_\text{t}^2$, which thereby allows for a topological transition~\cite{sato2:09,bernevig:13}.

The topological character of the SC state is given by the TKNN number
\begin{align}
C=\frac{\gamma}{2\pi}=\frac{1}{2\pi N}\sum_{\k}\Omega(\k),
\label{c7}
\end{align}
where
\begin{align}
\Omega(\k)=i\sum_{n,\lambda}\,\left[\bm\nabla_\k\times(u^*_{n\lambda}(\k)\bm\nabla_\k u_{n\lambda}(\k))\right]_{z}
\label{c6}
\end{align}
is the $z$-component of the Berry curvature~\cite{thouless:82}.
The sum over $n$ runs over the occupied bands $E_{\k,n}<0$ and $\lambda=1,\dots,4$ labels the components of the eigenvectors $\b u_n(\k)$ of the matrix $H_\k$.
The number $C$ is integer valued, if the occupied energy levels are separated by a finite gap from the unoccupied levels. The value of $C$ and therefore the topology of the quantum state changes when the energy gap closes at $|H_z|=H_\text{t}$. For magnetic fields $|H_z|>H_\text{t}$, the energy gap opens again (see discussion in Sec.~\ref{sec3}). This reopening of a gap, above the paramagnetic limiting field $\mu_\text{B}|H_z|=\Delta$, is tied to the presence of SOC, which protects the spin-singlet pairing channel~\cite{note1}.

\begin{figure}[t!]
\centering
\vspace{3mm}
\begin{overpic}
[width=1\columnwidth]{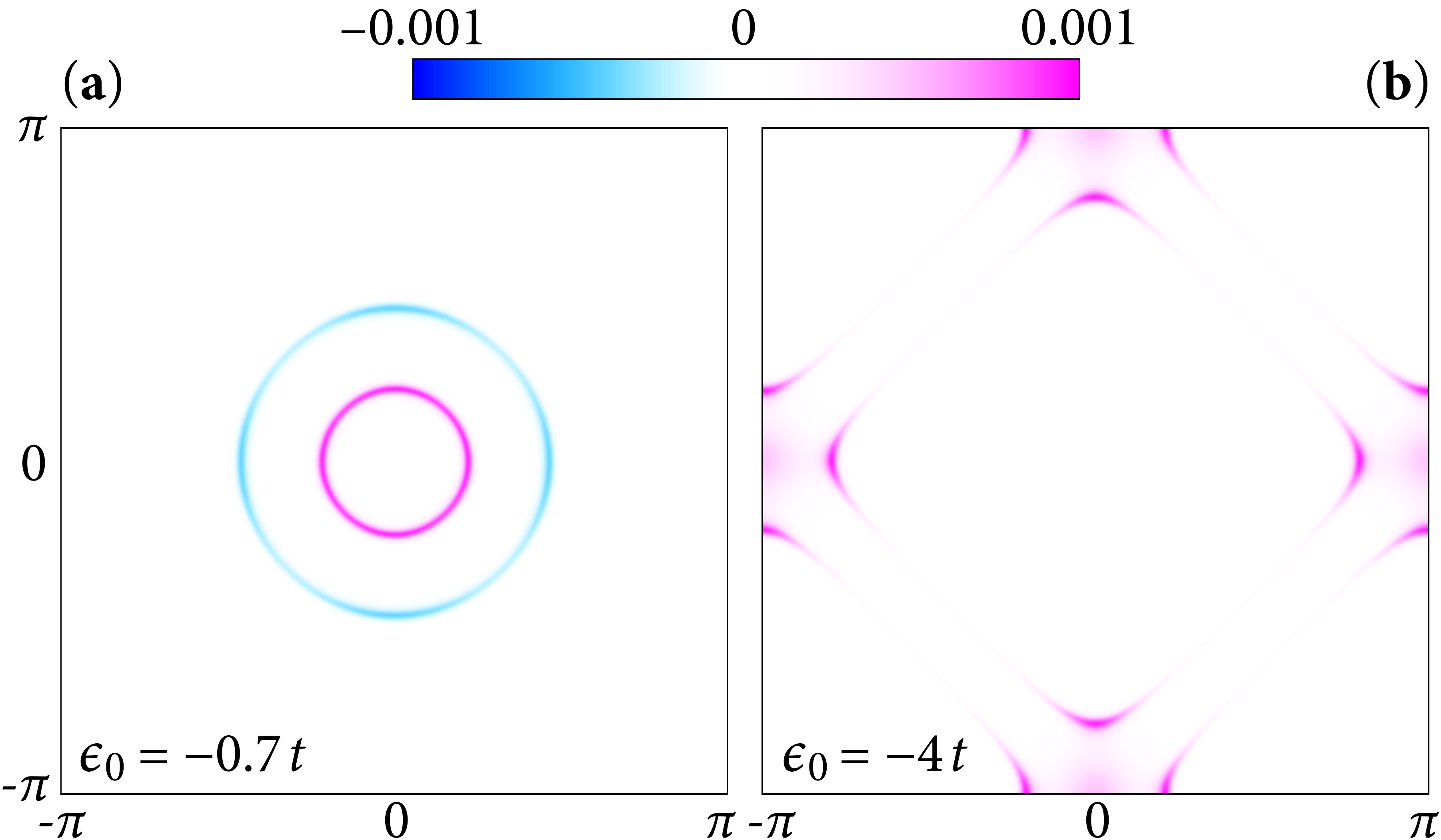}
%\put(5.5,45.5){(a)}
%\put(54.5,45.5){(b)}
\end{overpic}
\vspace{-2mm}
\caption{$z$-component of the Berry curvature $\Omega(\k)$ for an out-of-plane magnetic field $H_z$ in the SC state. $\Omega(\k)$ is finite within the window $\Delta$ below the Fermi energy. (a) In the topologically trivial state, (here: $\epsilon_0=-0.7\,t$, $\mu_\text{B}H_z=0.3\,t$ and $\Delta$ is fixed to $0.1\,t$), the total Berry curvature integrates to zero over the Brillouin zone. (b) In the topological situation (B) (see main text, $\epsilon_0=-4\,t$ and $\mu_\text{B}H_z=0.3\,t$), the Berry curvature integrates to $2\pi C=4\pi$ over the Brillouin zone.
}
\label{Fig2}
\end{figure}

Only the momenta $\k$ for which $\xi^\pm_\k$ lie within a window $\Delta$ below the Fermi energy contribute to $C$. The sign of this contribution reflects the winding direction of the $x$--$y$ components of the spin $(\hbar/2)\langle c^\dag_{\k,s}\bm\sigma_{ss'}c_{\k,s'}\rangle$ in momentum space (see Fig.~\ref{Fig1}). If $|H_z|<H_\text{t}$, the $\k$-integrated Berry curvatures in the vicinity of the two normal-state Fermi surfaces cancel exactly [Fig.~\ref{Fig2}~(a)] and consequently $C=0$.

The topological state emerging for $|H_z|>H_\text{t}$ is of different nature in two distinct density regimes:

\noindent (A) small electron density ($\mu<-2t$): 
the condition $|H_z|>H_\text{t}$ leads to $\xi^+_\k>\Delta$ for all $\k$ and therefore the $\xi^+_\k$-band is empty and does not contribute, i.e., the pink (positive) contributions to $C$ in Fig.~\ref{Fig2}~(a) vanish. Consequently, the superconducting state is characterized by $C=\mp1$, depending on the sign of $H_z$. This situation is realized for small band fillings.

\noindent (B) densities near half-filling ($|\mu|<2t$): in this regime, two separate topological transitions are possible.
At a magnetic field $\mu_\text{B}|H_z|=\sqrt{(\epsilon_0+4t)^2+\Delta^2}<\mu_\text{B}H_\text{t}$, the character of the $\xi^+_\k$-band changes from particle- to hole-like and thereby reverses the sign of $\Omega(\k)$ in the vicinity of the corresponding Fermi surface [Fig.~\ref{Fig2}~(b)]. Therefore, a topological transition to $C=\pm2$ occurs, depending on the sign of $H_z$, with both, the $\xi^-_\k$- and the $\xi^+_\k$-band, partially occupied. A realization of this superconducting state close to half filling is however unlikely due to possibly competing orders. For an even larger magnetic field $\mu_\text{B}|H_z|>H_\text{t}$, the $\xi_\k^+$-band is again lost, and $C$ changes to $=\pm1$.
The topological properties of both cases, (A) and (B), correspond to those described in Ref.~[\onlinecite{sato1:09}] in the context of spin-triplet superconductors.

A special characteristic feature of $s$-wave superconductivity in the presence of Rashba SOC is that the magnetic field induces an inter-band pairing contribution where a quasi-particle of the $\xi^+_\k$-band is paired with one from the $\xi^-_{\k}$-band at opposite momentum. This pairing contribution induces interior energy gaps above and below the Fermi energy (clearly visible e.g. in Fig.~\ref{Fig5}~(a) below, cf. Ref.~[\onlinecite{liu03}]). A more detailed discussion of the relation between intra- and inter-band pairing is given in appendix~\ref{app1}.

\subsection{In-plane magnetic field}
\label{sec2}

For a finite in-plane magnetic-field component $\b H_\parallel$, the inversion symmetry $\k\rightarrow-\k$ is broken, since $\xi^\pm_\k\neq\xi^\pm_{-\k}$, and the Fermi surfaces are shifted out of the Brillouin-zone center in opposite directions perpendicular to $\b H_\parallel$ [Fig.~\ref{Fig1}~(c)]. The pairing of electrons with momenta $\k$ and $-\k$ is thereby suppressed. Instead, pairs are formed in which electrons have momenta $\k$ and $-\k+\q^\pm$, respectively, where the COMMs $\q^\pm$ account for the Fermi surface shifts~\cite{barzykin:02,kaur:05,michaeli:12,loder:13} (see appendix~\ref{app2}).

The SC ground state with an in-plane magnetic field component therefore contains in general two order parameters $\Delta_{\q^+}$ and $\Delta_{\q^-}$. These enter the generalized on-site pairing term as
\begin{align}
{\cal H}_{\rm I}=\sum_{\k,\q}\left[\Delta_\q^{\!*}c_{-\k+\q,\da}c_{\k,\ua}+\Delta_\q c^\dag_{\k,\ua}c^\dag_{-\k+\q,\da}\right],
\label{g3}
\end{align}
where $\q=\q^+,\q^-$~\cite{loder:10}. The singlet order parameter for COMM $\q$ is calculated self-consistently from
\begin{align}
\Delta_\q=-\frac{V}{2N}\sum_{\k'}\langle c_{-\k'+\q,\da}c_{\k',\ua}-c_{-\k'+\q,\ua} c_{\k',\da}\rangle,
\label{g4}
\end{align}
where $V$ is the strength of the pairing-interaction. With increasing in-plane magnetic-field strength $|\b H_\parallel|$, the difference $|\q^+-\q^-|$ grows. Such a finite COMM state is spatially non-uniform~\cite{loder:10} with lines of zero pair density, similar to the SC state proposed by Larkin and Ovchinnikov for a singlet superconductor in a strong Zeeman field~\cite{larkin:64}. Characteristic for this state is a mixing of intra- and inter-band pairing and the absence of a full energy gap (see appendix~\ref{app1} and \ref{app3})~\cite{loder:10,loder:13}.

\begin{figure}[t!]
\centering
\vspace{3mm}
\begin{overpic}
[width=1\columnwidth]{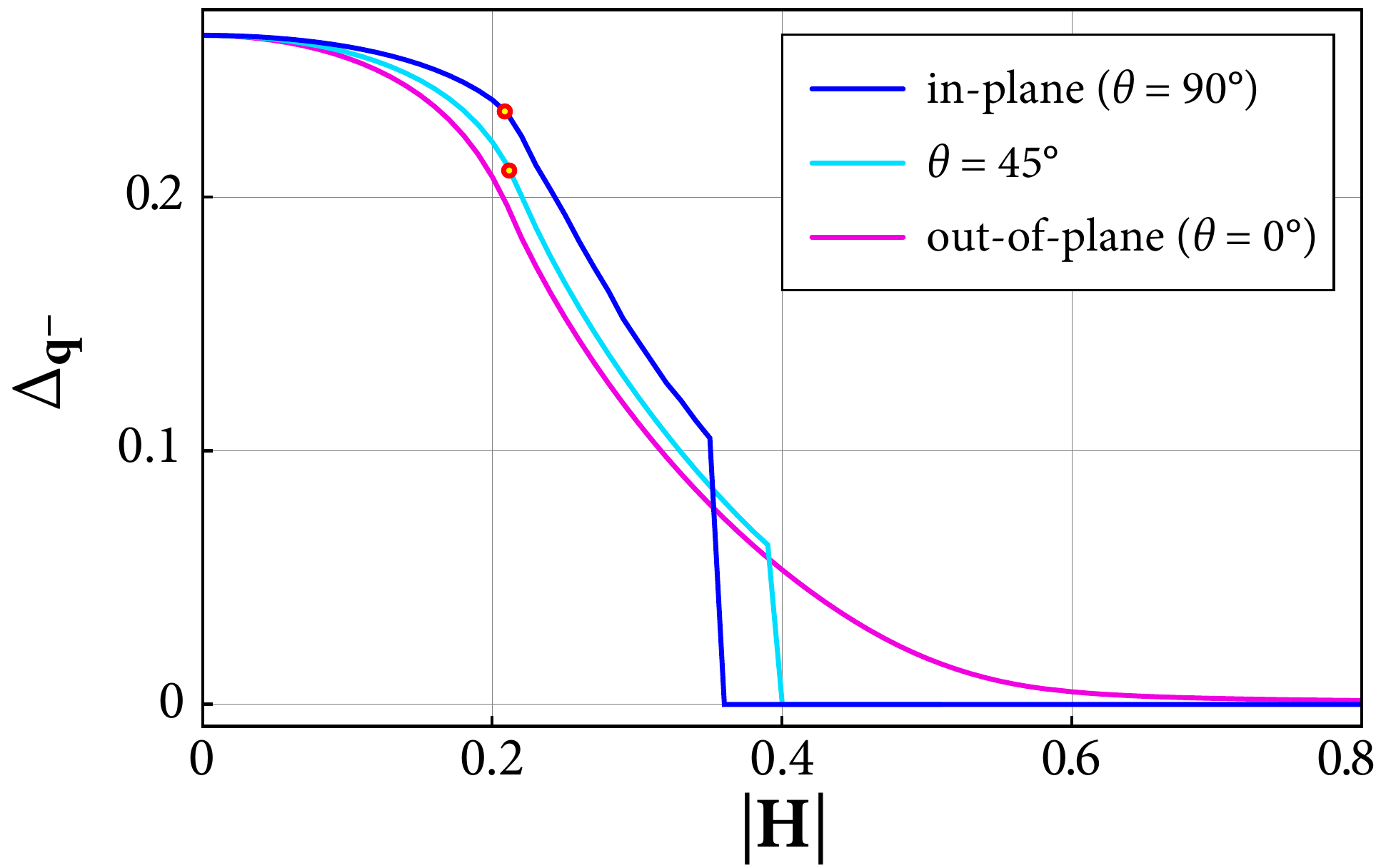}
\end{overpic}
\vspace{-2mm}
\caption{Self-consistent solutions of the SC order parameter $\Delta_{\q^-}$ for three magnetic-field directions and $V=4\,t$, $\alpha=0.5\,t$ and a constant electron density $n=0.05$. This value of $n$ corresponds to $\epsilon_0\approx-0.17\,t$. For such low densities, a large interaction strength $V$ is required to obtain a reasonably large order parameter. For each value of $|\b H|$, $\q^-=(q,0)$ is obtained by minimizing the free energy. The red circles indicate the magnetic field strength above which a finite COMM $q\neq0$ is present.
}
\label{Fig3}
\end{figure}

For $|\b H_\parallel|\neq0$, the topological characterization is more involved. In situation (B), close to half-filling, two COMMs $\q^\pm$ appear and the absence of a full energy gap implies that $C$ is not an integer and therefore unsuitable to characterize the topology (nevertheless, edge modes may still occur, see appendix~\ref{app3}). For this reason, we focus below on situation (A) when only the lower band $\xi^-_\k$ is occupied and therefore only electron pairs with COMM $\q^-$ form. Such a state is spatially uniform and similar to the state introduced by Fulde and Ferrell~\cite{fulde:64}, but it carries a finite charge current perpendicular to $\b H_\parallel$. If also $H_z\neq0$, it exhibits a full energy gap and therefore $C$ is integer-valued, except for a limited crossover region $H_\text{t}<|\b H|<H_\text{t}^+$ discussed below. Equation~(\ref{g6}) and therefore the definition of $H_\text{t}$ is valid as well in the presence of an in-plane magnetic field component whereby $H_z$ is replaced by $|\b H|$.

If $|\b H|>H_\text{t}$, one finds that $\alpha|\b g_\k|<\mu_\text{B}|\b H|$ for all occupied momenta $\k$ in the $\xi_\k^-$-band. Furthermore, if $\b H$ is strictly in-plane, say along the $y$-axis, ${\b h}_\k$ is parallel to $\b H$ on the $k_x$-axis and therefore the spins are parallel ($\xi^-_\k$-band) or anti-parallel ($\xi_\k^+$-band) to $\b H$ as well [Fig.~\ref{Fig1}~(c)]. As a consequence, no intra-band pairing in the singlet channel is possible for $k_y=0$, i.e., the intra-band energy gap closes at the two Fermi points with $k_y=0$. This gap closing for an in-plane field orientation corresponds to a topological transition from $C=-1$ to $C=1$ (cf. Fig.~\ref{Fig4}).

The topological phase of situation (A) with an in-plane magnetic field component is described by the same Hamiltonian as in Eq.~(\ref{g5}) replacing $\Delta$ by $\Delta_{\q^-}$.
In Sec.~\ref{sec3} we show that the topological state found for a sufficiently strong out-of-plane magnetic field can persist when the field is rotated --- even down to an in-plane field orientation with arbitrarily small $H_z$.

\section{Phase diagram and edge states}
\label{sec3}

\begin{figure}[t!]
\centering
\vspace{3mm}
\begin{overpic}
[width=0.9\columnwidth]{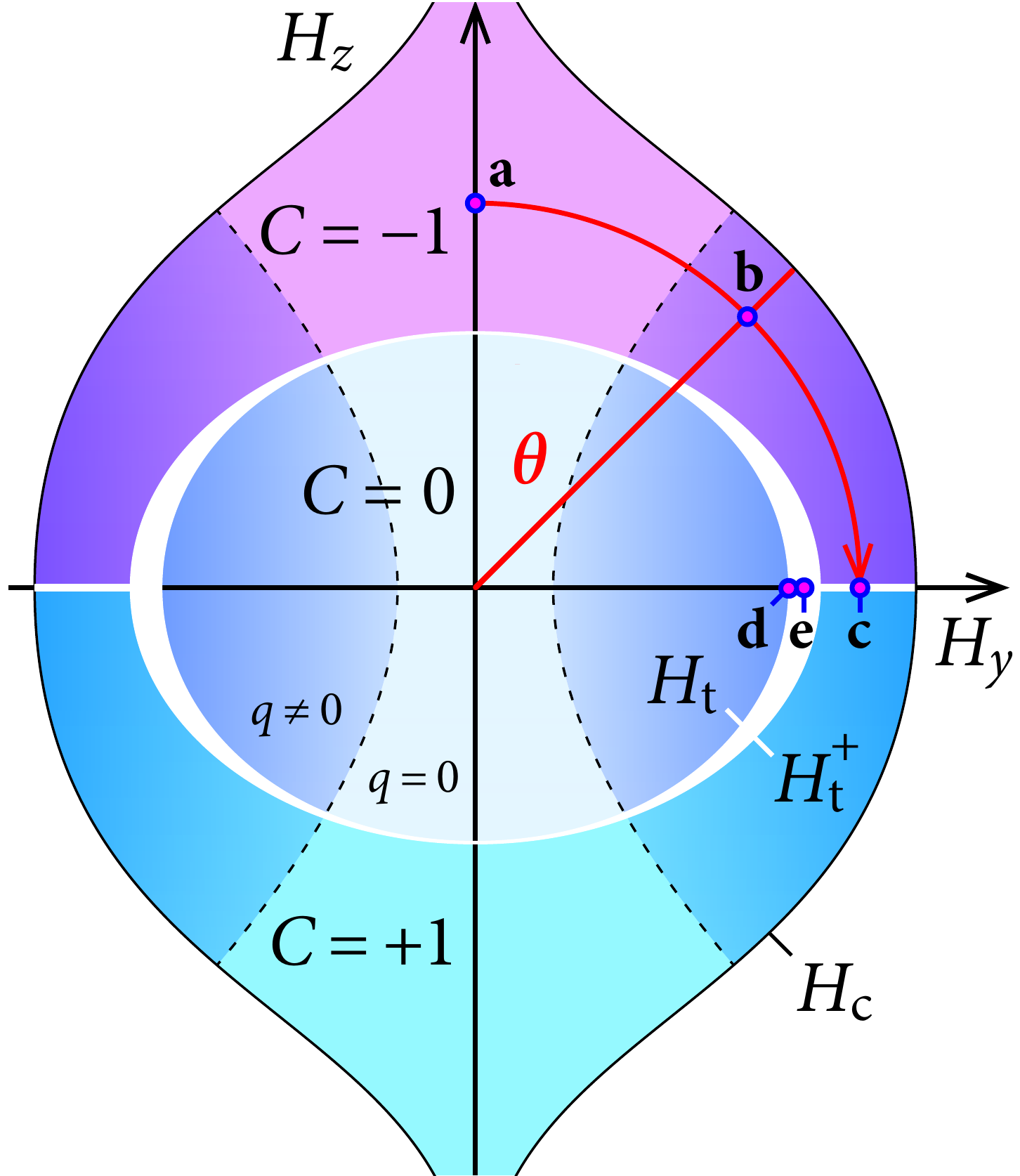}
\end{overpic}
\vspace{-2mm}
\caption{Phase diagram showing the topologically different SC states as a function of out-of-plane magnetic field $H_z$ and in-plane magnetic field $H_y$, not including orbital coupling to the magnetic field. The blue circles (a--e) mark the $H_y$--$H_z$-points for which the energy spectra are shown in Fig.~\ref{Fig5}. The dashed lines indicate the transition from zero COMM to finite COMM pairing.
}
\label{Fig4}
\end{figure}

\begin{figure*}[t!]
\centering
\vspace{3mm}
\begin{overpic}
[width=2.07\columnwidth]{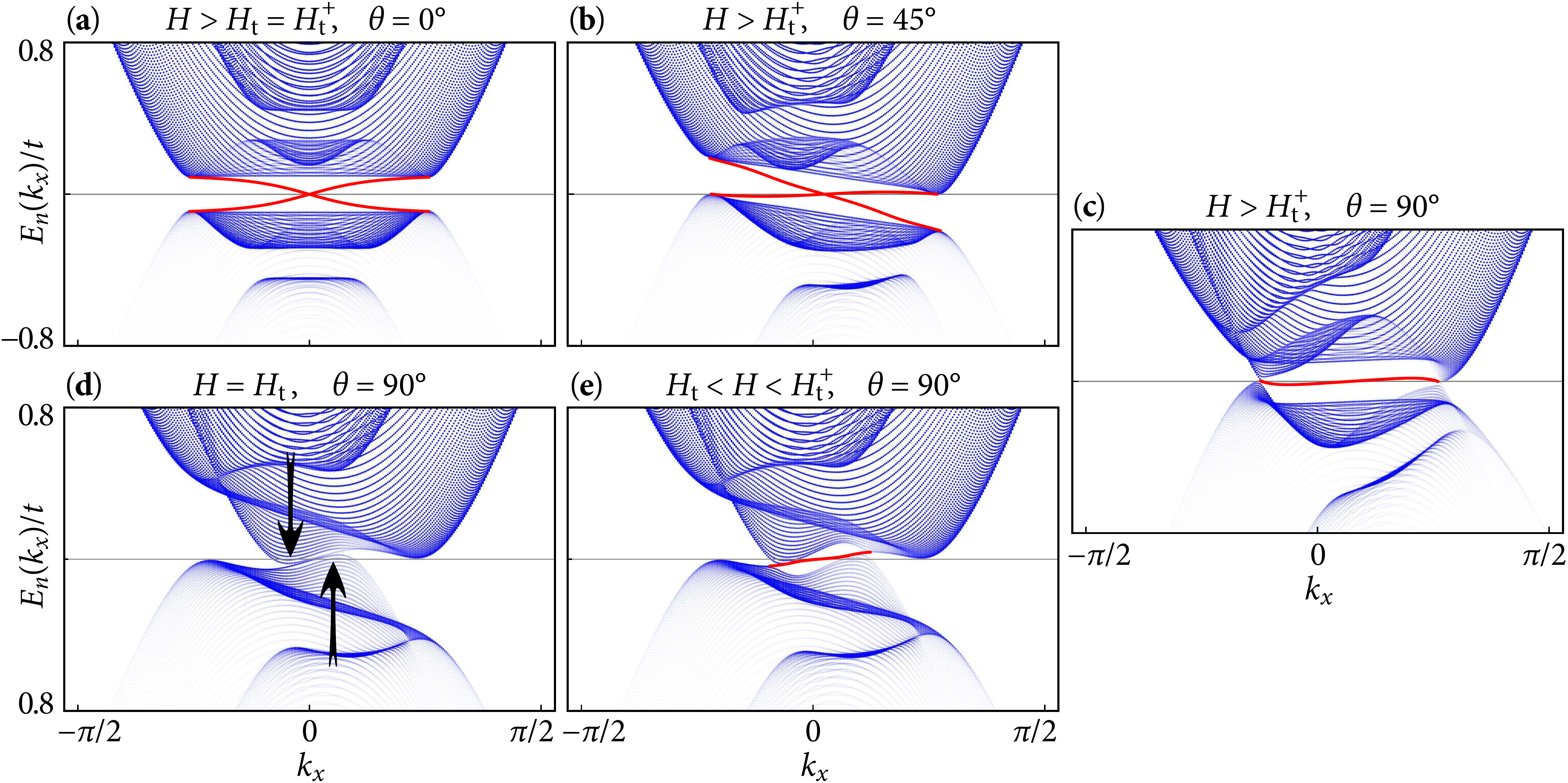}
%\put(4.6,42){(a)}
%\put(36.7,42){(b)}
%\put(68.9,32){(c)}
%\put(4.6,21.5){(d)}
%\put(36.7,21.5){(e)}
\end{overpic}
\vspace{-2mm}
\caption{Energy spectra $E_n(k_x)$ for a stripe geometry with 600$\times$100 sites, open boundary conditions and in-plane magnetic field component in $y$-direction, and parameters $V$, $\alpha$, and $n$ as in Fig.~\ref{Fig3}. (a--c) The evolution of the edge modes (red lines) upon rotating the magnetic field is shown for (a) $\theta=0$, (b) $\theta=45^\circ$, (c) $\theta=90^\circ$, and $\mu_\text{B}|\b H|=0.3\,t$. The self-consistently calculated order parameters $\Delta_{\q^-}$ and COMMs $\q^-$ are (a) $\Delta_{(0,0)}=0.11\,t$, (b) $\Delta_{(0.02\pi,0)}=0.12\,t$, and (c) $\Delta_{(0.14\pi,0)}=0.14\,t$. (d,e) illustrate the crossover regime $H_\text{t}(90^\circ)\leq H_y\leq H_\text{c}(90^\circ)$: (d) $\mu_\text{B}H_y=\mu_\text{B}H_\text{t}(90^\circ)=0.23\,t$ and $\Delta_{(0.12\pi,0)}=0.21\,t$, and (e) $\mu_\text{B}H_y=0.24\,t <\mu_\text{B}H^+_\text{t}(90^\circ)$ and $\Delta_{(0.12\pi,0)}=0.20\,t$. The black arrows in (d) indicate the partial occupation of states originating from the $\xi^+_\k$-band. The opacity of each point encodes the weight with which the corresponding state contributes to the density of states.
}
\label{Fig5}
\end{figure*}

We start the analysis of the SC state with the discussion of the self-consistent solutions of the SC order parameter. Figure~\ref{Fig3} shows the magnetic-field dependence of $\Delta_{\q^-}$ for different angles $\theta$ of the field direction. The Rashba SOC ensures the presence of a finite in-plane spin component which allows for singlet pairing. Therefore the Zeeman coupling to a field in $z$-direction ($\theta=0^\circ$ and $\q^-=\bm0$) cannot wipe out superconductivity completely (pink curve) when orbital depairing is not included. A finite in-plane field component leads instead to a finite critical magnetic field $H_\text{c}(\theta)$, above which  there are no solutions for $\Delta_{\q^-}$.

An interesting result is the somewhat larger value for $\Delta_\q$ in an in-plane magnetic field $|\b H_\parallel|<H_\text{c}(\theta)$ than in an out-of-plane magnetic field of the same magnitude. Consequently, the magnetic field $H_\text{t}(\theta)$ at which the energy gap closes, grows with increasing angle $\theta$ and is maximal for an in-plane direction. Likewise, the field $H_\text{t}^+(\theta)$ above which the energy gap opens again and the topological state emerges, is maximal for $\theta=90^\circ$, whereas $H_\text{t}^+(0^\circ)=H_\text{t}(0^\circ)$. The resulting phase diagram for different magnetic-field orientations is qualitatively drawn in Fig.~\ref{Fig4}. The topologically trivial SC state ($C=0$) is bounded by the ellipse given by $H_\text{t}(\theta)$, which itself is within the slightly larger ellipse formed by $H_\text{t}^+(\theta)$. The white regime in between separates the state with $C=0$ from the states with $C=\pm1$. In this crossover region the energy gap is closed and $C$ is not an integer. A further topological transition occurs for the in-plane field orientation $\theta=90^\circ$ and $|\b H|>H_\text{t}^+$: if $\theta$ sweeps through 90$^\circ$, the out-of-plane field component $H_z$ changes sign and, accordingly, $C$ changes from $-1$ to 1. As discussed above, the energy gap is closed as well along this transition line.

The importance of finite-COMM pairing for the topological properties of the SC state is illustrated using the energy spectra shown in Fig.~\ref{Fig5}. These spectra are calculated for a stripe geometry with open boundary conditions in $y$-direction, which allows for in-gap edge modes (drawn in red). We choose the in-plane magnetic-field component in the $y$-direction and thereby obtain a  shift of the Fermi surfaces out of the Brillouin-zone center in $k_x$-direction. Therefore a COMM $\q^-=(q,0)$ ($q\geq0$ for $H_y>0$) has to be taken into account for pairing in the $\xi_\k^-$-band. The free energy of the SC state is minimized for the smallest $q$ which is still large enough to avoid an indirect closing of the energy gap (see Fig.~\ref{Fig5}~(a--c) and appendix~\ref{app2}). The TKNN number $C$ thereby remains well defined up to the magnetic-field direction $\theta=90^\circ$. For $\theta\rightarrow90^\circ$ [Figs.~\ref{Fig5}~(c)], the energy gap closes at two $k_x$-points, and a topological transition occurs.
The magnetic-field strength, above which a finite $\q^-$ is required (dashed line in Fig.~\ref{Fig4}), depends on the angle $\theta$, but it is necessarily smaller than $H_\text{t}$, if the field direction is mainly in-plane.

Eventually, Fig.~\ref{Fig5} also shows the evolution of the in-gap edge modes under the rotation of the magnetic field~\cite{note2}.
Starting from $\theta=0^\circ$, the energy difference $|\xi^-_\k-\xi^-_{-\k}|$ grows for increasing $\theta$ and thus the dispersion of the two opposite edge modes becomes asymmetric [Fig.~\ref{Fig5}~(b)]. Upon approaching $\theta=90^\circ$, the energy gap closes at two $k_x$ points [Fig.~\ref{Fig5}~(c)]. Consequently, $q$ must be chosen to ensure that these closing points are located at the Fermi energy in order to prevent the energy bands above and below from overlapping. In this situation, the two edge modes are degenerate. Indeed, the two modes carry edge currents flowing in the same direction opposite to the flow of the bulk current. These modes are similar to the edge modes found for $p$-wave superconductors in an in-plane magnetic field~\cite{wong:13,schnyder:11,queiroz:14}, except for the presence of a finite COMM pairing due to the Rashba SOC.

Figures~\ref{Fig5}~(d,e) illustrate the gap closing and the emergence of edge modes in the regime $H_\text{t}(90^\circ)\leq H_y\leq H_\text{t}^+(90^\circ)$. At $H_y=H_\text{t}(90^\circ)$, the energy gap closes at $k_x=0$ [Fig.~\ref{Fig5}~(d)]. However, the minimum of the $\xi_\k^+$-band is at a momentum $k_x<0$ and somewhat below the Fermi energy (indicated by the left black arrow), whereas the maximum of its mirrored hole-band is at a momentum $k_x>0$ somewhat above the Fermi energy (right black arrow). Thus, the $\xi_\k^+$-band and its mirrored hole band $-\xi^+_{-\k+\q^-}$ overlap indirectly. Superconductivity nevertheless persists because of  the gain of condensation energy from the $\xi_\k^-$-band. In the regime $H_\text{t}(90^\circ)\leq H_y\leq H_\text{t}^+(90^\circ)$, a direct gap opens again around $k_x=0$ with two gap-crossing edge modes, although the indirect gap remains closed [Fig.~\ref{Fig5}~(e)].

For magnetic fields $H_y>H_\text{t}^+$, all states of the $\xi_\k^+$-band are above the Fermi energy. In this regime an infinitesimally small out-of-plane magnetic-field component $H_z$ is sufficient to remove the two gap-closing points and ensure well defined TKNN numbers $C=\pm1$. Eventually, superconductivity breaks down at $H_y=H_\text{c}(\theta)$: Above this critical magnetic field, the two gap-closing points move into the continuum of the energy bands above and below the Fermi energy. The upper and lower bands therefore overlap and the self-consistent solution for the SC order parameter is lost. Although $H^+_\text{t}(\theta)$ depends only weakly on the SOC strength $\alpha$, the critical magnetic field $H_\text{c}(\theta)$ grows with increasing $\alpha$. In order to obtain $H_\text{c}(90^\circ)>H_\text{t}^+(90^\circ)$, it is required that $\alpha>\mu_\text{B}H_y$.

\section{Discussion}
\label{sec4}

How can the above topologically non-trivial SC state be realized in a solid-state system? We infer that an ideal candidate system would consist of several partially filled energy bands with a sizable Rashba SOC. Such a model~\cite{joshua:12,zhong:13} was proposed e.g. to describe the physics of the conducting interface between LaAlO$_3$ and SrTiO$_3$~\cite{ohtomo:04,thiel:06}.

The condition $\mu_\text{B}|\b H|>\sqrt{\epsilon_0^2+\Delta^2}$ for $C\neq0$ implies that the lower limit for the Zeeman splitting is given by $\Delta$. The corresponding magnetic field is typically larger than the upper critical field $H_{\text{c}2}$ above which orbital depairing destroys superconductivity. Therefore the topological state is not accessible with a magnetic field oriented along the $z$-axis. The topological state can be reached only for a nearly in-plane field orientation with $H_z<H_\text{c2}$ but $|\b H|>H_\text t^+$. This excludes the situation (B) with a close to half-filled band, because of the presence of two different COMMs $\q^\pm$ as discussed above. The alternative situation, on the other hand, requires  that $\mu_\text{B}|\b H|>|\epsilon_0|$, which is close to the Fermi energy $E_\text{F}$ in a one-band model. For a band filling large enough to allow for a SC state (for a reasonable interaction strength $V$), $E_\text{F}$ for this partially filled band must be at least several meV. The magnetic field required to overcome this energy would be far too large for experimental realizations.

\begin{figure}[t!]
\centering
\vspace{3mm}
\begin{overpic}
[width=0.8\columnwidth]{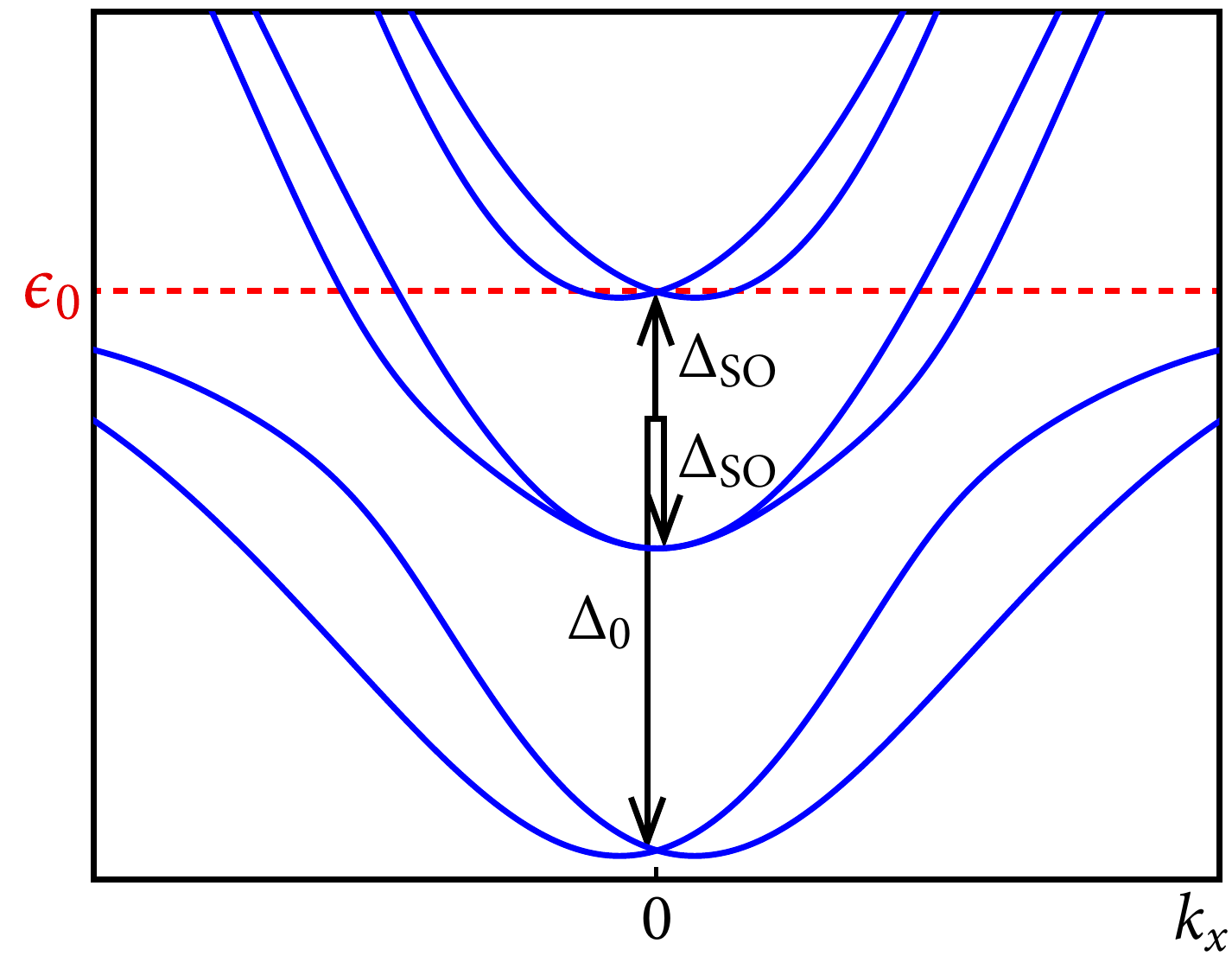}
\end{overpic}
\vspace{-2mm}
\caption{Band structure of the three-band model for $k_y=0$. In order to ensure $\epsilon_0=0$ (red dashed line), the Fermi energy should be at the degeneracy point of the upper, Rashba-like doublet. The parameters are here: $t_y^{xz}/t^{xz}_x=0.1$, $\Delta_0=t$, $\Delta_\text{SO}=\Delta_z=0.2\,t$.
}
\label{Fig6}
\end{figure}

In a multi-band setup, $\epsilon_0$ refers to the energy of the degeneracy point of a spin-orbit coupled doublet hosting the possibly topological state (see Fig.~\ref{Fig6}), relative to $E_\text{F}$.
The necessity of stabilizing a superconducting state with $\epsilon_0\rightarrow0$ requires the presence of at least two bands at $E_\text{F}$:
A lower band provides the electron density for a sufficient gain of condensation energy in the superconducting state, whereas a second band has a minimum close enough to the Fermi energy so that it can be emptied or partially filled through an external control parameter.

A model which implements these features, and additionally also a strong Rashba-like SOC, recently emerged from the theoretical description of the LAO-STO interface. At this interface, the intrinsic electrostatic potential in LAO induces a nearly 2D electron liquid, which resides mainly in the titanium $3d$ $t_{2g}$ orbitals of the first TiO$_2$ layer~\cite{thiel:06,breitschaft:10}. The band-structure of this oxide interface provides a prototype for a class of similar interface systems that we introduce here by constructing a tight-binding Hamiltonian ${\cal H}={\cal H}^0+{\cal H}^\text{SO}+{\cal H}^z$ for the three $t_{2g}$ bands $d_{xy}$, $d_{xz}$, and $d_{yz}$, following Refs.~[\onlinecite{joshua:12,zhong:13}]. The free kinetic energy is given by
\begin{align}
{\cal H}^0=\sum_\k{\bf C}^\dag_\k\left[\begin{pmatrix}\epsilon^{xz}_\k&0&0\cr0&\epsilon^{yz}_\k&0\cr0&0&\epsilon^{xy}_\k-\Delta_0\end{pmatrix}\!\otimes\sigma^0\right]{\bf C}_\k,
\label{9}
\end{align}
where $\sigma^0$ is the 2$\times$2 unity matrix and ${\bf C}_\k^\top=(c^{xz}_{\k\ua},c^{xz}_{\k\da},c^{yz}_{\k\ua},c^{yz}_{\k\da},c^{xy}_{\k\ua},c^{xy}_{\k\da})$. The hopping matrix elements $t^{xy}_x=t^{xy}_y$ for the $d_{xy}$ band are identical in the $x$- and $y$-direction, whereas they are different for the $d_{xz}$ and $d_{yz}$ bands: $t^{xz}_x=t^{yz}_y\simeq t^{xy}_x\gg t^{xz}_y=t^{yz}_x$. Furthermore, the in-plane $d_{xy}$ orbital is lowered in energy by $\Delta_0$ relative to the out-of-plane $d_{xz}$ and $d_{yz}$ orbitals because of the symmetry breaking interface.

The spin-orbit coupling on the Ti atoms is described by
${\cal H}^\text{SO}=\Delta_\text{SO}\sum_\k{\bf C}^\dag_\k[{\bf L}\otimes\bm\sigma]{\bf C}_\k$, where the angular momentum operator $\bf L$ for $l=2$ is represented in the $\{d_{xz}$,\,$d_{yz}$,\,$d_{xy}\}$ basis~\cite{joshua:12}. This term intermixes the $t_{2g}$ orbitals and generates three doublets; the upper two doublets are split by $2\Delta_\text{SO}$ (see Fig.~\ref{Fig6}). In addition, the deformation of the $t_{2g}$ orbitals due to the interface potential leads to a hybridization of the $d_{xz}$, $d_{yz}$ orbitals with the $d_{xy}$ orbital, parameterized by
\begin{align}
{\cal H}^z=i\Delta_z\sum_\k{\bf C}^\dag_\k\!\left[\!\begin{pmatrix}0&0&\!-\sin k_x\!\cr0&0&\!\sin k_y\cr \sin k_x&-\sin k_y&0\end{pmatrix}\!\otimes\sigma^0\right]\!{\bf C}_\k.
\label{13}
\end{align}
The $\k$-dependence in ${\cal H}^z$ splits the three otherwise doubly degenerate doublets. For small momenta $k_x$ and $k_y$, this splitting acts on the lowest and the highest doublet exactly like a Rashba term in a one-band model. This source of an effective Rashba-like band splitting can be several orders of magnitude larger than the splitting through the relativistic term and thereby is able to explain qualitatively the spin splitting observed at the LAO-STO interface~\cite{caviglia:10,fete:12}.
Further, if the Fermi energy is tuned to $\Delta_0+\Delta_\text{SO}$ by an external gate voltage~\cite{caviglia:10}, $\epsilon_0\simeq0$ is fulfilled for the highest doublet (see Fig.~\ref{Fig6}). The SC state can be stabilized by the two lower doublets, whereas the highest generates a non-trivial topological number $C=\pm1$.

This three-band model is likely the minimal model which fulfils the requirements discussed above for the realization of topological $s$-wave superconductivity in a solid. Various interface systems with a similar setup are conceivable,
however, the formation of a topological SC state is viable only in a restricted parameter range: The Rashba-like band splitting, which is controlled by the parameter $\alpha$ in the one-band model, is replaced in the three-band model by $\alpha_\text{R}=2\Delta_\text{SO}\Delta_z/\Delta_0$~\cite{zhong:13}. Although the magnetic field $H_\text{t}^+$, above which $C\neq0$ is possible, varies little with $\alpha_\text{R}$, the magnetic field range $H_\text{c}-H_\text{t}^+$ vanishes when $\alpha_\text{R}$ approaches zero. To ensure a wide magnetic field range for the topological state, the parameter $\alpha_\text{R}$ should therefore be larger than the Zeeman splitting and thus also larger than the SC energy gap.

In the following we estimate that the above criteria are indeed satisfied in the candidate system LAO-STO. The interface superconducts below a critical temperature of about 300\,mK~\cite{reyren:07,caviglia:08} and exhibits an energy gap $\Delta$ of about 40\,$\upmu$eV with most likely $s$-wave symmetry~\cite{richter:13}. The Rashba parameter $\alpha_\text{R}$ was experimentally estimated to be in the range 20--100\,meV~\cite{caviglia:10}, which is compatible with $\alpha_\text{R}$ determined from the three-band model using $\Delta_0$, $\Delta_\text{SO}$, and $\Delta_z$ from the band-structure calculations of Ref.~[\onlinecite{zhong:13}]. Assuming that $\epsilon_0$ can be adjusted to zero by a suitable gate voltage, the necessary Zeeman splitting $\mu_\text{B}|\b H|>\Delta$ is far smaller than $\alpha_\text{R}$ and corresponds to a magnetic field $H_\text{t}\approx600$\,mT (the in-plane $H_\text{t}$ might be somewhat larger). While the measured out-of-plane critical field is $H_\text{c}(0)\approx200$\,mT~\cite{reyren:07} and therefore smaller than $H_\text{t}$, the observed in-plane critical field is $H_\text{c}(90^\circ)\gtrsim1\,\text{T}>H_\text{t}$~\cite{reyren:09}.

The other important parameter defining $H_\text{t}$ is $\epsilon_0$, which is controlled by the electron density $n$ at the interface. The electron density can be tuned between 1$\times$$10^{-13}\,\text{cm}^{-1}$ and 6$\times$$10^{-13}\,\text{cm}^{-1}$~[\onlinecite{caviglia:08,breitschaft:10,richter:13}]. Using $\Delta_0\approx50$\,meV as in Ref.~[\onlinecite{joshua:12}] and $\Delta_\text{SO}\approx20$\,meV~\cite{zhong:13}, we find that setting $\epsilon_0=0$, i.e. $E_\text{F}=\Delta_\text{0}+\Delta_\text{SO}$ (see Fig.~\ref{Fig6}), requires $n\approx3\times10^{-13}\,\text{cm}^{-1}$. This lies well within reach by a gate voltage. A precise prediction for the electron density at which the topological state should first develop is however difficult, since the value of $\Delta_0$ and the position of the higher bands is under debate. Density-functional calculations provide rather a value $\Delta_0\approx250$\,meV~\cite{zhong:13}. However, to account for the low electron densities measured experimentally, the $d_{xy}$ electrons are likely to be localized in this scenario. The Fermi energy is in this case measured relative to the lower edge of the $d_{xz}$, $d_{yz}$ orbitals, i.e., $\epsilon_0=0$ for $E_\text{F}=2\Delta_\text{SO}$. This is realizable as well within the range of charge densities tunable through a gate voltage.

The verification of a parameter regime which allows for a topological superconducting state could come from measuring the Knight shift in the nuclear magnetic resonance (NMR) frequency, e.g. of the La-nuclei in the first LAO layer. Due to the band splitting generated through the Rashba-type SOC, the superconductor magnetizes in a magnetic field, with a spin susceptibility $\chi_\text S$ remaining finite down to $T=0$. Therefore, if a Rashba SOC with strength $\alpha_\text{R}\gg\Delta$ is indeed present at the LAO-STO interface, the drop of the Knight shift at $T_\text{c}$ must be far smaller than expected for a standard $s$-wave superconductor. Further, a kink in $\chi_\text S$ upon changing the gate voltage might reveal the voltage at which the highest doublet starts to get occupied and the search for topological edge states is most promising.

\acknowledgements{The authors gratefully acknowledge discussions with Hans Boschker, Daniel Braak, Peter Fulde, Jochen Mannhart, Christoph Richter, Jonathan Ruhman and Kevin Steffen. This work was supported by the DFG through TRR 80.}

\clearpage
\appendix

\section{Order Parameter in Spin- and Band-Space}
\label{app1}

Of particular interest is the relation of the spin-singlet order parameter $\Delta$ to the intra- and inter-band pairing amplitudes in terms of the eigenstate operators $a_{\k\pm}$ of the SOC split bands. 
In spin space, the spin-singlet order parameter is represented, as in Eq.~(5) of the main text, by the 2$\times$2 matrix
\begin{align}
\hat\Delta=i\sigma^y\Delta=\begin{pmatrix}0&\Delta\cr-\Delta&0\end{pmatrix}
\label{a1}
\end{align}
where 
\begin{align}
\Delta=\frac{V}{2N}\sum_\k\langle c_{\k,\da}c_{-\k,\ua}-c_{\k,\ua}c_{-\k,\da}\rangle.
\label{a2}
\end{align}
In this section we focus exclusively on pairing with zero center-of-mass momentum (COMM), although the following derivations are valid for arbitrary magnetic field directions. Finite COMM as required in the case of a sufficiently large in-plane component of the magnetic field (see Sec.~\ref{app2}) is implemented simply by replacing $-\k$ by $-\k+\q$. 

\begin{figure}[t!]
\centering
\vspace{3mm}
\begin{overpic}
[width=0.99\columnwidth]{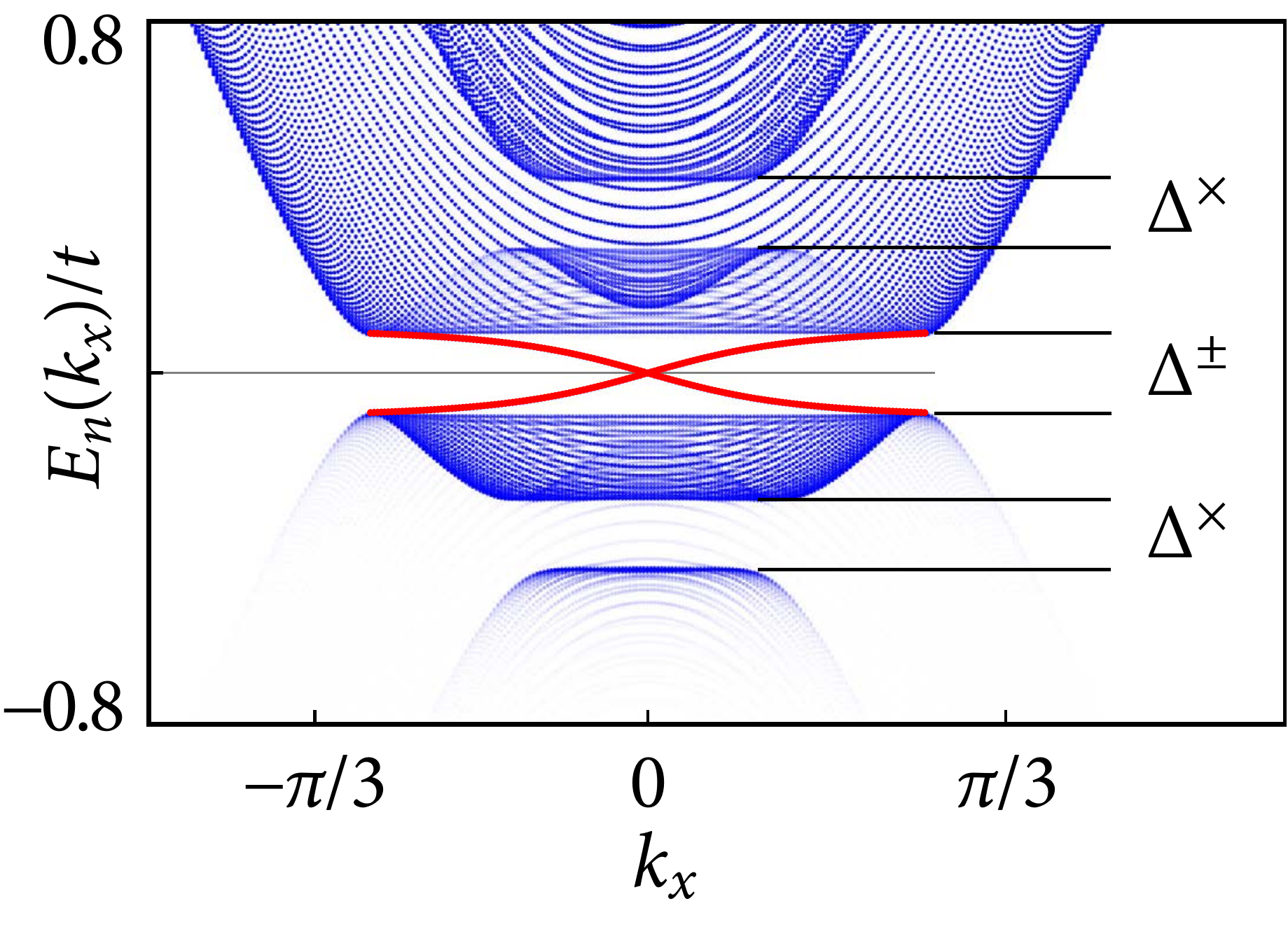}
%\put(-6,65.5){(c)}
\end{overpic}
\vspace{-2mm}
\caption{\small{Illustration of the energy gaps induced by the different pairing contributions of an $s$-wave superconductor with an out-of-plane magnetic field $H_z>H_\text{t}$ (with parameters as in Fig.~5 of the main text). The energy gap centered around $E=0$ originated from the intra-band pairing term $\Delta^+=\Delta^-$, whereas the inter-band pairing term $\Delta^\times$ generates two interior gaps above and below the Fermi energy, where the normal-state bands $\xi^+_\k$ and $-\xi^-_{-\k}$, or similarly $\xi^-_\k$ and $-\xi^+_{-\k}$, cross.}
}
\label{SFig1}
\end{figure}

In the presence of SOC, the operator $c_{\k s}$ no longer annihilates an eigenstate of the normal state Hamiltonian. Instead, the band operators $a_{\k\pm}$ refering to the eigenstates with energies $\xi^\pm_\k$ are obtained by the transformation 
\begin{align}
\begin{pmatrix}a_{\k+}\cr a_{\k-}\end{pmatrix}={\frac{{h_{\k,x}-ih_{\k,y}}}{2h_\k}}\begin{pmatrix}1&\phi_\k^+\cr1&\phi_\k^-\end{pmatrix}\begin{pmatrix}c_{\k,\ua}\cr c_{\k,\da}\end{pmatrix}
\label{a3}
\end{align}
with
\begin{align}
\phi^\pm_\k=\frac{h_{\k,z}\pm h_\k}{h_{\k,x}-ih_{\k,y}},
\label{a4}
\end{align}
and $h(\k)=|\b h(\k)|$ is the absolute value of the Bloch vector $\b h(\k)$. Applying this transformation on $\hat\Delta$ generates an order parameter $\tilde\Delta$ for pairing the normal state eigenstates with energies $\xi^\pm_\k$:
\begin{align}
\tilde\Delta_\k&=\begin{pmatrix}\Delta_\k^+&\Delta_\k^\times\vspace{2mm}\cr-\Delta_\k^\times&\Delta_\k^-\end{pmatrix}\nonumber\\
&=\frac{{h_{\k,x}^2+h_{\k,y}^2}}{4h_\k^2}\begin{pmatrix}-(\phi_\k^+-\phi_{-\k}^+)\Delta&-(\phi_\k^+-\phi_{-\k}^-)\Delta\vspace{2mm}\cr(\phi_\k^+-\phi_{-\k}^-)\Delta&-(\phi_\k^--\phi_{-\k}^-)\Delta\end{pmatrix}.
\label{a5}
\end{align}
Here, $\Delta^+_\k=-\Delta^-_\k$ represents the intra-band pairing gaps, whereas $\Delta^\times_\k$ represents the inter-band pairing gap. In zero magnetic field, $\phi_\k^+-\phi_{-\k}^-=0$ and therefore only intra-band pairing occurs. However, in finite magnetic fields there is always a finite inter-band order parameter $\Delta^\times_\k$. This term opens a gap at the intersections of the dispersions $\xi^+_\k$ and $-\xi^-_{-\k}$, or $\xi^-_\k$ and $-\xi^+_{-\k}$, which are above and below the Fermi energy, respectively (illustrated in Fig.~\ref{SFig1}). These interior gaps are enforced by the assumed contact pairing interaction, which allows exclusively for $s$-wave singlet pairing. Inter-band pairing and interior gaps are likely absent for spin-triplet pairing. Note that for an out-of-plane magnetic field, the absolute values $|\Delta^\pm_\k|=\Delta^\pm$ and $|\Delta^\times_\k|=\Delta^\times$ are $\k$-independent. In the presence of an in-plane field component, also the absolute values are $\k$-dependent.

\begin{figure*}[t!]
\centering
\vspace{3mm}
\begin{overpic}
[width=0.99\columnwidth]{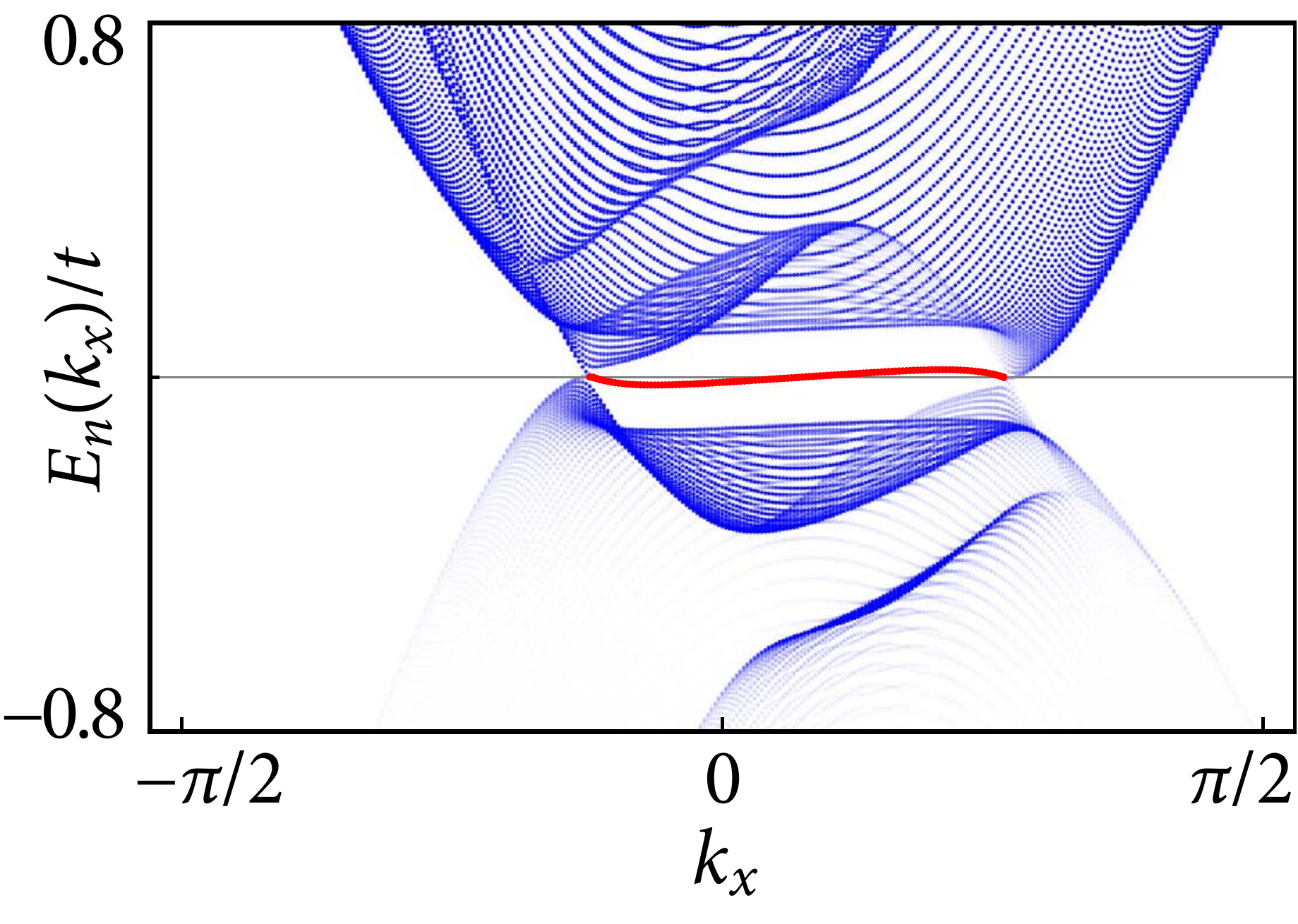}
\put(-4,66.5){\large\bf a)}
\end{overpic}\hspace{5mm}
\begin{overpic}
[width=0.99\columnwidth]{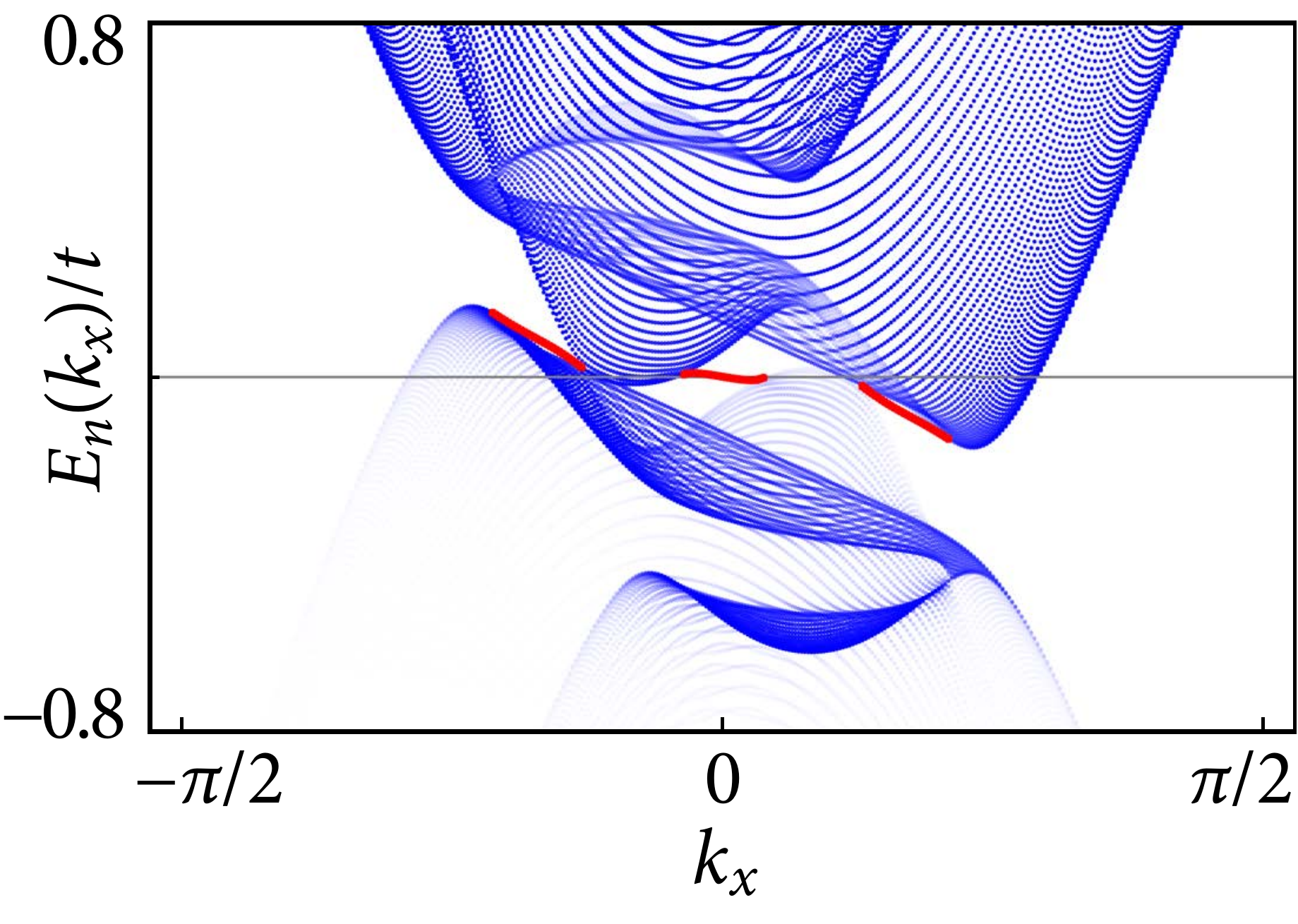}
\put(-4,66.5){\large\bf b)}
\end{overpic}
\vspace{-2mm}
\caption{\small{Illustration of the effect of a finite COMM on the energy gap of an $s$-wave superconductor in an in-plane magnetic field. The inversion asymmetry $\xi^\pm_\k\neq\xi^\pm_{-\k}$ in the simultaneous presence of an in-plane magnetic field and Rashba SOC leads to tilted edges of the $\xi^-_\k$-continuum above and below the energy gap (b). This tilting can be corrected by choosing an appropriate COMM $\q^-$ which compensates the shift of the energy bands. (a) redisplays Fig.~5~(c) in the main text, whereas in (b), $\Delta$ is fixed to the same value as in (a), but with $\q^-=0$. In (b), two degenerate edge modes are present as well, although they merge with the continuum around the two direct-gap closing points.}
}
\label{SFig3}
\end{figure*}

If $\b H$ is orientated strictly in-plane (here we assume $\b H=\b H_\parallel=(0,H_y,0)$), then
\begin{align}
\phi^+_\k=-\phi^-_\k=\frac{h_{\k,x}-ih_{\k,y}}{h_\k}.
\label{a7}
\end{align}
It follows that in this case, 
\begin{align}
\phi_\k^\pm-\phi_{-\k}^\pm=\left\{\begin{array}{ll}1,&\alpha|\sin k_x|>\mu_\text BH_y\\0,&\alpha|\sin k_x|<\mu_\text BH_y\end{array}\right.
\end{align}
for all $\k$ on the $k_x$-axis. Since in the regime $H_y>H_\text{t}$, $\mu_\text BH_y>\alpha|\sin k_{\text F\!,\,x}|$ is always fulfilled, the intra-band gaps $\Delta_\k^\pm$ close on the two Fermi points $k^\pm_{\text F\!,\,x}$ on the positive and the negative $k_x$-axis, respectively, for $H_y>H_\text{t}$ [see Fig.~\ref{SFig3}~(a)]. This gap closing demonstrates the impossibility to form spin-singlet pairs in the intra-band channel on the $k_x$-axis in the regime $\alpha|\sin k_y|<\mu_\text BH_y$ (cf. the spin configuration on the $k_x$-axis in Fig.~1~(c)).

\section{In-plane magnetic fields and finite momentum pairing}
\label{app2}

In this section we illustrate the necessity of pairing with a finite COMM $\q$ in the simultaneous presence of a Rashba SOC and an in-plane magnetic field. Figure~\ref{SFig3}~(a) is identical to Fig.~5~(c) of the main text, showing the energy spectrum in a stripe geometry for an in-plane magnetic field $H_y>H_\text{t}$. Since in the normal state of this \textquotedblleft situation (A)\textquotedblright\ only the $\xi^-_\k$-band is partially occupied, naturally only intra-band pairing occurs, with a COMM $\q^-$. Because the intra-band energy gap appears at the crossings of the two bands $\xi^-_\k$ and $-\xi^-_{-\k+\q^-}$, the ideal situation for pairing would be the existence of a $\q^-$ for which $\xi^\pm_\k=\xi^\pm_{-\k+\q^-}$ for all $\k$. This condition cannot be fulfilled exactly for a tight-binding description (since the in-plane magnetic field slightly deforms the band structure). Nevertheless it is sufficient for the formation of electron pairs, if $\xi^\pm_\k-\xi^\pm_{-\k+\q^-}<\Delta$ for all $\k$ on the respective Fermi surface. The COMM $\q^-$ chosen to fulfil this latter condition ensures that the indirect energy gap does not close [cf. Fig.~\ref{SFig3}~(a)], except at the two $\k$-points described in the main text.

In fact, the free energy is minimized by the COMM $\q^-=(q,0)$ [for $\b H_\parallel=(0,H_y,0)$] with the smallest $q$ for which the energy gap does not close indirectly. This condition translates into the requirement that the energy difference
\begin{align}
\delta\xi(q)=\xi^-_{\k_\text{F}^+}-\xi^-_{-\k_\text{F}^++\q^-}
\label{c7}
\end{align}
must be smaller than the energy gap around the Fermi energy. Here, $\k_\text{F}^+$ is the Fermi momentum on the positive $k_x$ axis. If the magnetic field orientation is not close to in-plane, i.e., $H_y<\alpha\sin k^\pm_{\text F\!,\,x}$, the energy gap is simply given by $\Delta$. However, upon approaching an in-plane orientation, the energy gap closes directly at the two Fermi points on the $k_x$-axis. Therefore, $q$ is fixed by the condition $\delta\xi(q)=0$ for an in-plane magnetic field.

\begin{figure*}[t!]
\centering
\vspace{3mm}
\begin{overpic}
[width=0.99\columnwidth]{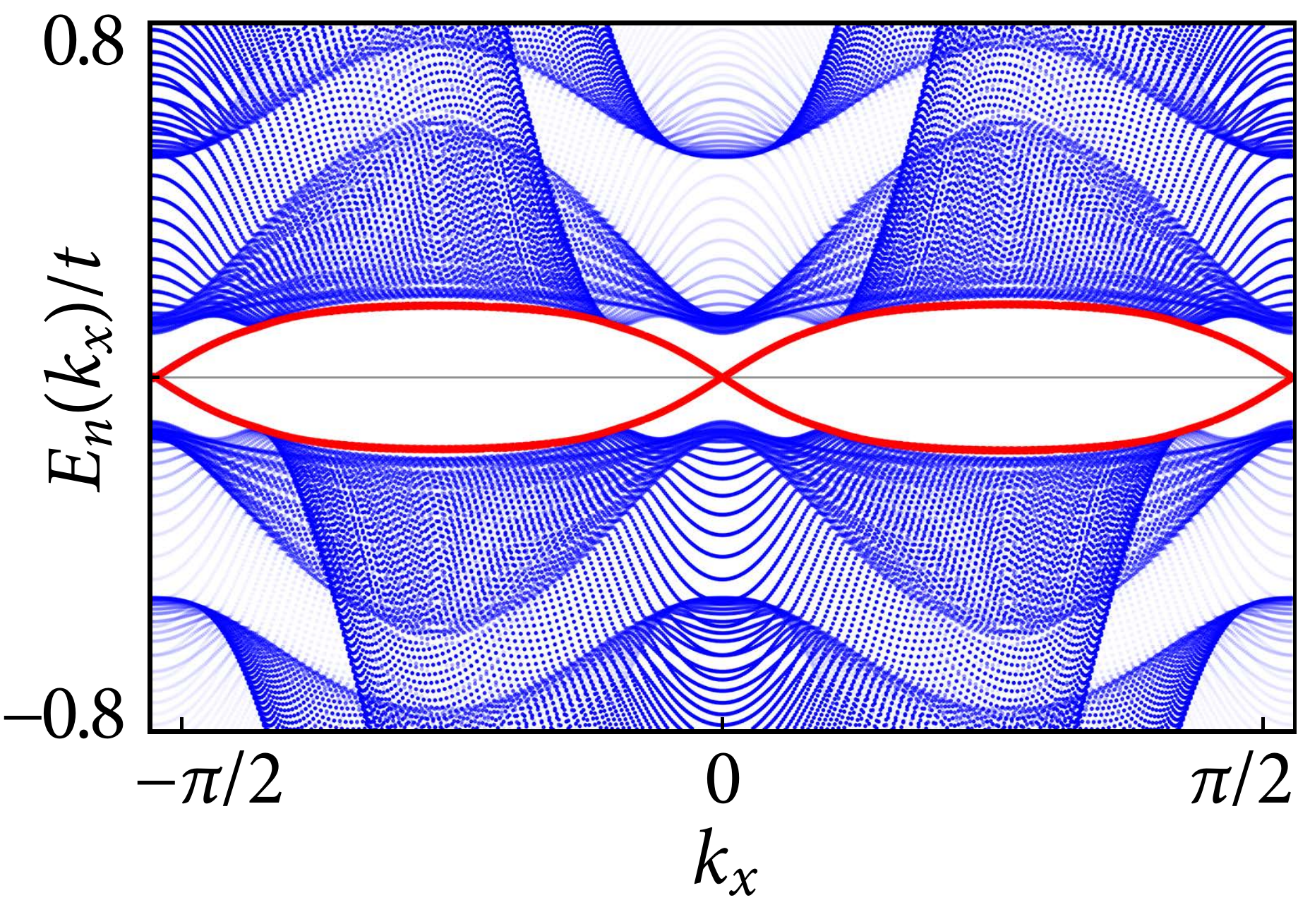}
\put(-4,66.5){\large\bf a)}
\end{overpic}\hspace{5mm}
\begin{overpic}
[width=0.99\columnwidth]{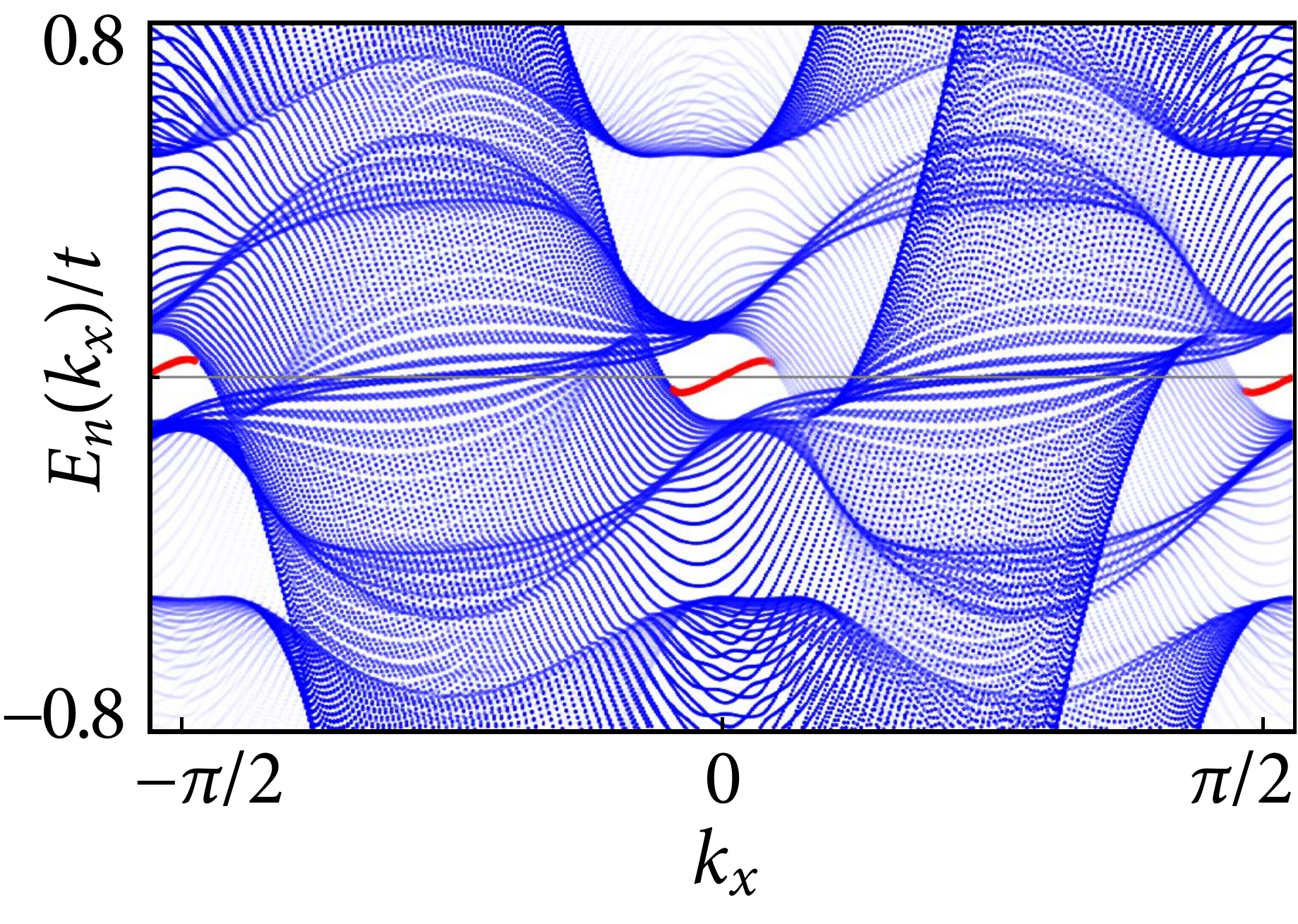}
\put(-4,66.5){\large\bf b)}
\end{overpic}\\[3mm]
\begin{overpic}
[width=0.99\columnwidth]{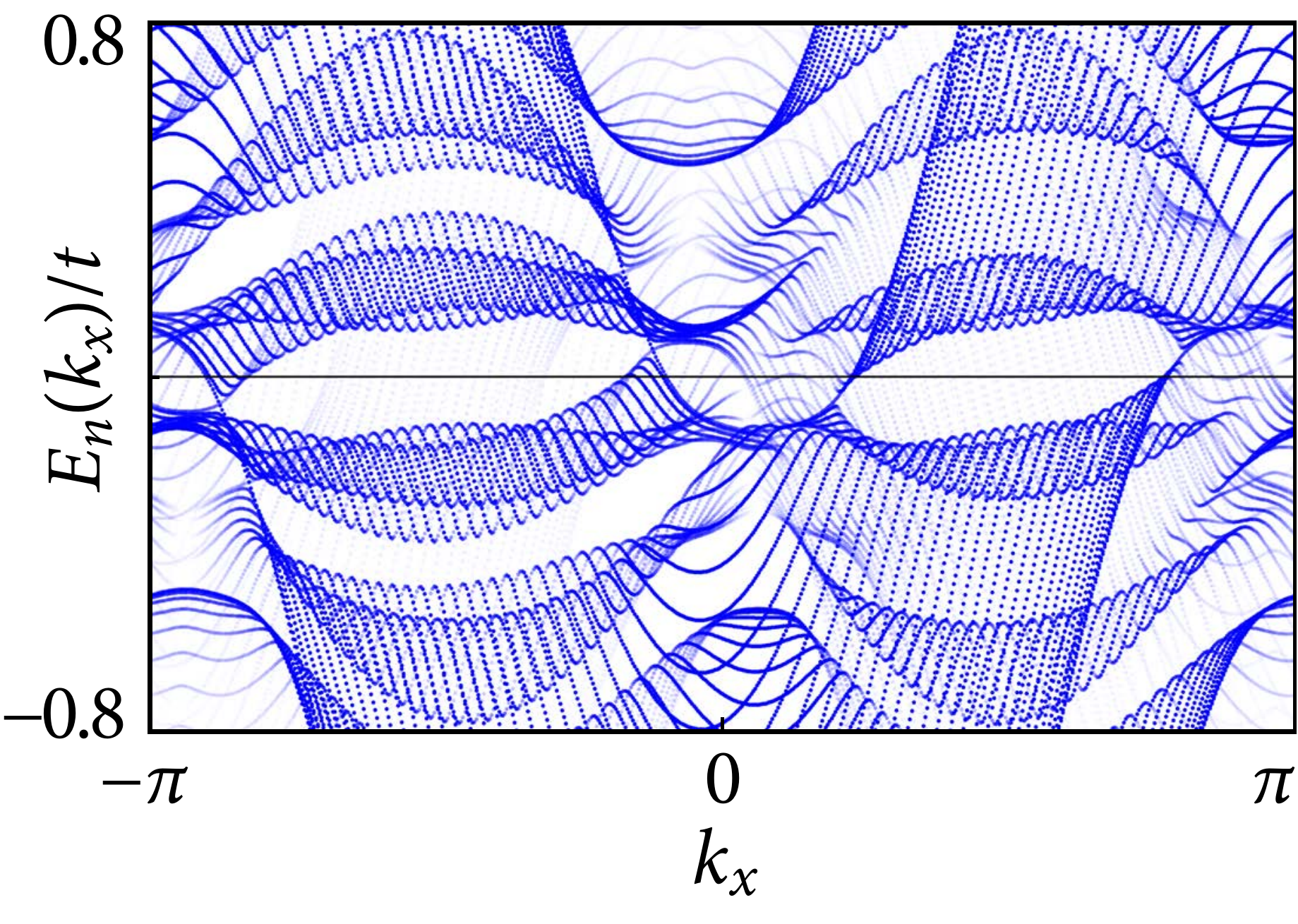}
\put(-4,66.5){\large\bf c)}
\end{overpic}
\vspace{-2mm}
\caption{\small{Energy spectrum of the topologically non-trivial state $\mu_\text B|\b H|>\sqrt{(\epsilon_0+4t)^2+\Delta^2}$ for \textquotedblleft situation B\textquotedblright\ with $\epsilon_0+4t=\mu=0$, $\mu_\text B|\b H|=0.3\,t$ and (a) an out-of-plane magnetic field, (b) an in-plane magnetic field with zero COMM and (c) an in-plane magnetic field with COMMs $\q^+$ and $\q^-$ optimizing pairing in the $\xi^+_\k$- and the $\xi^-_\k$-band. In (a), there is a full energy gap with topological edge states crossing each other twice, i.e., $C=2$. While no energy gap around the Fermi energy remains in (b), finite COMM pairing can only partially recover the energy gap. The spectra in (a) and (b) are calculated for a stripe geometry. The spectrum in (c) is calculated using periodic boundary conditions, therefore edge states are absent.}
}
\label{SFig2}
\end{figure*}

Figure~\ref{SFig3}~(b) shows the energy spectrum of the same superconducting system as Fig.~\ref{SFig3}~(a), but setting $q$ ad hoc to zero. The energy gap in this case is obviously indirectly closed. Additionally, two direct closing points exist at momenta $|k^\pm_x|<|k_{\text F\!,\,x}^\pm|$.
It is important to notice that the self-consistent solution for $\Delta$ vanishes abruptly, if the energy gap closes indirectly upon increasing $\b H$. In Fig.~\ref{SFig3}~(b), $\Delta$ is kept fixed for illustration purposes, although no self-consistent solution of this kind exists.

\section{Topology in $\bm S$-Wave Superconductors close to Half-Filling}
\label{app3}

In the main text we argued that in the density regime close to half filling ($|\mu|<2t$, entitled \textquotedblleft situation (B)\textquotedblright), no integer topological number $C$ can be defined, because finite COMM pairing with $\q^+$ and $\q^-$ necessarily generates in-gap states. This phenomenon is specifically illustrated here in Fig.~\ref{SFig2}:
For an out-of-plane magnetic field, the topological state for $\mu_\text BH_z>\sqrt{(\epsilon_0+4t)^2+\Delta^2}$ has $C=2$~\cite{sato1:09}. Consequently, the emerging edge states cross each other twice, at $k_x=0$ and at $k_x=\pi$ [Fig.~\ref{SFig2}~(a)]. Such a topological state is possible for field orientations forming a small angle with the $z$-axis. Upon rotating the magnetic field into the plane, but keeping $\q^\pm=\bm0$, the edge states become degenerate but remain well defined in those sections of $k_x$ where a direct energy gap persists. [Fig.~\ref{SFig2}~(b)]. 
However, the shift of the two Fermi surfaces in opposite directions leads to a closing of the energy gap over an extended region in momentum space. As discussed above, such a situation with $\mu_\text BH_y>\sqrt{(\epsilon_0+4t)^2+\Delta^2}$ but $\q=\bm0$ does not correspond to a state of minimum free energy, but serves only for demonstrating the effect of the in-plane magnetic field.

Unlike in the low-density regime ($\mu<-2t$, \textquotedblleft situation (A)\textquotedblright), it is here not possible to remove the in-gap states by compensating the shift of the Fermi surfaces with finite COMMs. Since both, the $\xi^+_\k$- and the $\xi^-_\k$-band, are partially occupied, two order parameters $\Delta_{\q^+}$ and $\Delta_{\q^-}$ are required, where $\q^+\approx-\q^-$. The corresponding order parameter in real space exhibits a stripe-like pair-density-wave modulation of the form $\cos^2(q^+_x-q^-_x)$, which has lines of zero pair density~\cite{agterberg:08,berg09,loder:10}. This implies that the density of states of such a superconductor does not vanish at the Fermi energy, i.e., in-gap states necessarily remain.

The energy spectrum of this superconducting state is shown in Fig.~\ref{SFig2}~(c), where $\q^+$ and $\q^-$ are chosen to compensate the shift of the respective Fermi surfaces ideally. Unlike in Figs.~\ref{SFig2}~(a) and~(b), this spectrum is calculated using periodic boundary conditions, since the assignment of the COMMs $\q^+$ and $\q^-$ to the correct pairing terms in real space is numerically difficult. Therefore edge states are absent.


\begin{references}
\bibitem{note1}
In the presence of SOC, the spin susceptibility $\chi_\text{S}$ of the spin-singlet superconductor remains finite down to $T=0$. In particular, if $\Delta\ll\alpha$, $\chi_\text{S}$ is almost equal the Pauli susceptibility of the normal state. Therefore, the Clogston-Chandrasekhar paramagnetic limit $H_\text{CC}=\Delta/\sqrt{2}$~\cite{clogston:62} does also not apply.
\bibitem{note2}
For $\theta>0$, the direction of the in-plane field component relative to the boundary is important. If the in-plane field is orthogonal to the boundary (here: $y$-direction), the $k_x$-dispersion of the edge mode is unperturbed and remains gapless.
If there is a field component parallel to the boundary (in $x$-direction), the edge-mode disperses in $y$-direction as well. Since the dispersion in $y$-direction is quantized through the edges, the edge modes acquire  an energy gap around the Fermi energy (also observed for the edge modes in chiral $p$-wave superconductors~\cite{sato1:09,wong:13}). This mesoscopic energy gap is of the size of the normal-state level spacing around $E_\text F$ and vanishes like $1/M$ in the thermodynamic limit, where $M$ is the number of lattice sites in $y$-direction.
\end{references}
\end{document}